\documentclass[%
 reprint,
 amsmath,amssymb,
 aps,
]{revtex4-2}

\usepackage{graphicx}
\usepackage{dcolumn}
\usepackage{bm}
\usepackage{bbm}
\usepackage[colorlinks=true, allcolors=blue]{hyperref}

\newcommand{\dd}{\mathrm{d}}
\newcommand{\ii}{\mathrm{i}}

\newcommand{\GF}{G_{\mathrm{F}}}
\newcommand{\abs}[1]{\left\lvert#1\right\rvert}

\begin{document}

\title{Fast Neutrino-Flavor Conversion with Attenuation and Global Lepton Gradient}

\author{Masamichi Zaizen}
\affiliation{Department of Earth Science and Astronomy, The University of Tokyo, Tokyo 153-8902, Japan}

\author{Hiroki Nagakura}
\affiliation{Division of Science, National Astronomical Observatory of Japan, 2-21-1 Osawa, Mitaka, Tokyo 181-8588, Japan}

\date{\today}

\begin{abstract}
Fast neutrino-flavor conversion (FFC) can nontrivially alter neutrino radiation field in core-collapase supernovae (CCSN) and binary neutron-star merger (BNSM) remnants.
However, its interplay with global geometry remains poorly understood because microscopic flavor conversion scales are much shorter than global transport scales.
We perform global quantum kinetic neutrino transport simulations in spherical geometry with neutrino and matter backgrounds, using an attenuated oscillation Hamiltonian.
We find that steep radial lepton gradients can suppress FFC, whereas the suppression is highly sensitive to the adopted attenuation parameter.
This behavior is explained by an adiabatic condition: flavor coherence can grow sufficiently only while the flavor wave remains on the unstable branch in the local dispersion relation during propagation.
Background variation shifts the unstable branch, while attenuation lengthens the growth timescale, making the flavor coherence following more difficult.
We provide an approximate formula for the adiabaticity that can be used directly in CCSN and BNSM models developed by classical neutrino transport simulations.
Our results show that attenuation artificially leads to an overestimation of the impact of background variation and should therefore be applied with caution in global simulations of neutrino flavor conversion.
\end{abstract}

\maketitle


\section{Introduction}
Neutrino flavor conversion is a game-changing phenomenon in the dense neutrino gases \cite{Duan:2010,Richers:2022b,Tamborra:2021,Volpe:2024,Yamada:2024,Johns:2025,Raffelt:2025} in core-collapse supernovae (CCSNe) and binary neutron-star merger (BNSM) remnants.
In such extreme environments, neutrino-neutrino forward scattering potentially becomes dominant, and then the off-diagonal components in the flavor basis modify the dispersion relation governing flavor correlations between different flavors.
The nonlinear refractive effects can bring a nontrivial flavor conversion, sometimes called collective neutrino oscillations, and significantly alter the neutrino radiation field compared to the classical neutrino transport \cite{Wu:2017a,Li:2021,Fernandez:2022,Just:2022,Ehring:2023,Ehring:2023a,Ehring:2026,Nagakura:2024,Mori:2025,Wang:2025,Wang:2025a,Lund:2025,Qiu:2025,Qiu:2025a,Akaho:2025,Akaho:2026}.
Thereby, the quantum kinetics of neutrinos has been vigorously investigated to obtain more accurate modeling of CCSNe and BNSMs.

One of the currently attractive nature is fast neutrino-flavor instability (FFI) \cite{Sawyer:2005,Sawyer:2016}, which is induced by a crossing between the angular distributions in electron-neutrino lepton number (ELN) and heavy-leptonic number (XLN) \cite{Morinaga:2022,Dasgupta:2025a}.
The appearance of FFI has been widely surveyed in dynamical simulations of CCSNe \cite{Nagakura:2019,Nagakura:2021b,Morinaga:2020,DelfanAzari:2020,Abbar:2020,Glas:2020,Harada:2022,Akaho:2024a,Cornelius:2025a} and BNSMs \cite{Wu:2017,George:2020,Richers:2022a,Kawaguchi:2025a,Nagakura:2025a,Froustey:2026}.
Simultaneously, to capture the nonlinear behaviors, numerical simulations of fast neutrino-flavor conversion (FFC) triggered by FFI have been carried out in local spatial boxes \cite{Bhattacharyya:2021,Bhattacharyya:2022,Wu:2021,Zaizen:2023,Zaizen:2023a,Xiong:2023b,George:2024,Nagakura:2025,Liu:2025b}, and their asymptotic states have been clarified from the perspectives of stability and conservation laws.

Particularly, to investigate the neutrino transport within the astrophysical objects, quantum kinetic simulations under the global geometry have been directly solved \cite{Nagakura:2022a,Nagakura:2023,Nagakura:2023a,Nagakura:2023b,Nagakura:2023d,Shalgar:2023,Shalgar:2024,Xiong:2023,Xiong:2024,Urquilla:2026}.
Due to the disparity between the oscillation scales and the global scales in time and space, the attenuation technique was proposed to relax the computational complexity \cite{Nagakura:2022a}.
The attenuation to the oscillation Hamiltonian reduces the strength of the self-interactions and makes the nonlinear and smaller-scale flavor conversion tractable even within the global neutrino transport frameworks.
The global simulations have exhibited that the attenuation does not affect the overall structures averaged over the quasi-steady states and preserves the spatial properties, as verified by convergence tests.
Note that the convergence demands to keep the hierarchy among the physical scales, such as neutrino oscillations and collisions.

In the global neutrino transport, the background properties with which neutrinos interact vary with radius.
Neutrinos advect through (semi-)transparent regions after passing from the emission surface, 
and geometrical effects reduce the neutrino flux and also make the angular distributions forwardly peaked.
This alters their self-interactions, and the refractive effect they receive also changes with radius.
However, that is not the only background variation modifying the dispersion relation.
In a realistic environment, the background fluid has a radial profile, and the effects of the matter potential and collisions vary continuously along their propagation.

Bhattacharyya {\it et al.} \cite{Bhattacharyya:2025} showed that matter inhomogeneity can stabilize a system that is initially unstable to FFC.
Focusing on the comoving frame along neutrino trajectory, the spatially varying matter potential explicitly shifts the unstable branches and damps flavor coherence before reaching a linear saturation.
However, the approach was limited to a two-beam model within a local periodic box.
The suppression behaviors induced by matter gradient potentially appears even in a realistic global neutrino transport simulation.
There, neutrinos receive refractive effects from background matter (as charged leptons) and from neutrinos themselves (as neutral leptons) during their propagation, subsequently undergoing a nonlinear flavor conversion.

Our study provides a comprehensive study to clarify the impact of the background inhomogeneity on multi-angle FFC in the global spherical geometry with the Dirichlet boundary.
It is then necessary to adopt the attenuation on the oscillation Hamiltonian to treat the global neutrino transport simultaneously.
Although the prescription preserves the asymptotic states under the homogeneous matter background, the prospect could be violated in considering the inhomogeneity.
Hence, we also revisit the impact of the attenuation in more realistic environments compared with the previous work.
And as an analytical approach, we discuss the adiabaticity of flavor instability through the radial profile of the dispersion relation based on local linear stability analysis, not on the comoving frame.
Through the indicator, we can estimate whether a flavor wave can sufficiently grow during propagation in the presence of background variation.

This paper is structured as follows.
We begin by introducing our global simulation, stability analysis, and background models in Sec.\,\ref{Sec:2}.
Then, we demonstrate numerical simulations for models summarized in Table\,\ref{tab:models} in Sec.\,\ref{Sec:3A}.
We then analyze the behaviors using the linear stability analysis based on the dispersion relation approach in Sec.\,\ref{Sec:3B} and \ref{Sec:3C}.
Finally, further discussions and conclusions will be provided in Sec.\,\ref{Sec:4}.
We use natural units to set $\hbar = c = 1$ and the metric signature $\eta^{\mu\nu} = \mathrm{diag}(+,-,-,-)$ throughout this paper unless indicated otherwise.

\section{Method}\label{Sec:2}
\subsection{Global QKE in Nonlinear Regime}
Quantum kinetic equation (QKE) \cite{Sigl:1993} for neutrino density matrix $\rho_{\nu}$ in spherically symmetric geometry is given by
\begin{equation}
\begin{split}
    &\frac{\partial}{\partial t}\rho_{\nu} + \frac{1}{r^2}\frac{\partial}{\partial r}\left(r^2\cos\theta_{\nu}\,\rho_{\nu}\right) + \frac{1}{r}\frac{\partial}{\partial\cos\theta_{\nu}}\left(\sin^2\theta_{\nu}\,\rho_{\nu}\right) \\
    &\,\,\,\,\,\,= -\ii\xi\left[\mathcal{H}_{\mathrm{osc}},\rho_{\nu}\right].
\end{split}
\label{eq:QKE}
\end{equation}
The second and third terms in the left hand side represent advection in coordinate and momentum space, respectively.
In spherical geometry, momentum-space advection generates a radial gradient in the neutrino number density at larger radii.

The oscillation Hamiltonian in the right hand side consists of a vacuum term determined by the neutrino mass and refractive terms arising from lepton potentials composed of charged leptons and neutral leptons (neutrinos themselves):
\begin{equation}
    \mathcal{H}_{\mathrm{osc}} = U\frac{M^2}{2E_{\nu}}U^{\dagger} + v^{\mu}\Lambda_{\mu} + \sqrt{2}\GF\,v^{\mu}\int\dd\Gamma^{\prime}\,v_{\mu}^{\prime}\rho_{\nu}^{\prime}.
\end{equation}
$\Lambda_{\mu} = \sqrt{2}\GF\,\mathrm{diag}\left[\{j_{\mu}^{\ell}\}\right]$ corresponding to the number currents with a specified charged lepton $\ell$, and $v_\mu = (1, \boldsymbol{v})$ the four-vector denoting the propagation direction.
Under the spherical symmetry, $\boldsymbol{v}$ is simplified to $\cos\theta_{\nu}$.
The momentum space integration is $\int\dd\Gamma = \int\,E_{\nu}^2\dd E_{\nu}\dd\boldsymbol{v}/(2\pi)^3$ in the flavor-isospin convention where antineutrinos cover the negative occupation with the negative energy as $\bar{\rho}_{\nu}(E_{\nu}) = - \rho_{\nu}(-E_{\nu})$.
And in our study, the oscillation Hamiltonian is rescaled by an attenuation parameter $\xi$ to make the problem computationally tractable \cite{Nagakura:2022a}.
This attenuation to the oscillation scale is allowed only when it is sufficiently shorter than the other physical scales.
The asymptotic behavior is not altered by attenuation unless this condition is violated.

\subsection{Local Stability Analysis in Linear Regime}
The previous work\,\cite{Bhattacharyya:2025} employs the two beam model, where linear stability analysis can be conducted on the comoving frame along the neutrino beam.
Since the strategy allows the spatial variations of matter density to connect with the neutrino trajectory, the matter gradient can be directly included in the stability analysis.
On the other hand, our setup has a multi-beam structure in global geometry and cannot be simplified in the same ways.
Thus, we first perform local linear stability analysis at every spatial point, where the background lepton gradient is not included.
Then, the background density is assumed to be locally homogeneous and provides refractive effects on a flavor coherence.
Imposing the plane-wave ansatz on the flavor coherence, the impact can be described as the dispersion relation of the flavor wave.
And using the dispersion relation as a function of the radius, we evaluate whether the growth rate exceeds the phase shift rate due to the background variation.

The traditional linear stability analysis \cite{Izaguirre:2017}, not along the comoving frame, is given by
\begin{equation}
    \det\Pi^{\mu\nu}(k;r) = 0,
\end{equation}
where 
\begin{equation}
    \Pi^{\mu\nu}(k;r) = \eta^{\mu\nu} + \xi \sqrt{2}\GF\int\dd\Gamma\, (\rho_{ee}-\rho_{xx})\frac{v^{\mu}v^{\nu}}{v^{\sigma}k_{\sigma}+\xi\omega_V}.
    \label{eq:DR}
\end{equation}
Here, $\omega_V$ denotes a vacuum frequency and $\rho_{\alpha\alpha}$ the diagonal components (distribution functions) of density matrix at each radial point in the classical steady state $t_{\mathrm{cls}}$.
The neutrino angular distribution has radial dependence due to the momentum advection, and thereby the dispersion relation also depends strongly on the radius.

\begin{table}[t]
    \centering
    \begin{tabular}{cc|cc|cc}\hline\hline
         ~~$\lambda_0/\mu_0$~~ & ~~$m$~~ & ~~$\xi$~~ & ~~$\Delta R\mathrm{\,[km]}$~~ & ~~$N_r$~~ & ~~$t_{\mathrm{fin}}\mathrm{\,[\mu s]}$~~ \\\hline
         $1$ & 0  & $10^{-4}$ & $50$ & $24576$ & $500$ \\
         $1$ & -1 & $10^{-4}$ & $50$ & $24576$ & $500$ \\
         $1$ & -3 & $10^{-4}$ & $50$ & $24576$ & $500$ \\
         $4$ & 0  & $10^{-4}$ & $50$ & $24576$ & $500$ \\
         $4$ & -1 & $10^{-4}$ & $50$ & $24576$ & $500$ \\
         $4$ & -3 & $10^{-4}$ & $50$ & $24576$ & $500$ \\\hline
         
         $1$ & 0  & $4\times10^{-4}$ & $50$ & $98304$ & $500$ \\
         $1$ & -3 & $4\times10^{-4}$ & $50$ & $98304$ & $500$ \\
         $4$ & -3 & $4\times10^{-4}$ & $50$ & $98304$ & $500$ \\
         $10$ & -3 & $4\times10^{-4}$ & $50$ & $98304$ & $500$ \\
         $1$ & -3 & $4\times10^{-4}$ & $50$ & $24576$ & $500$ \\
         $4$ & -3 & $4\times10^{-4}$ & $50$ & $24576$ & $500$ \\\hline
         
         $1$ & 0 & $10^{-2}$ & $0.5$ & $24576$ & $10$ \\
         $1$ & -3 & $10^{-2}$ & $0.5$ & $24576$ & $10$ \\
         $20$ & -3 & $10^{-2}$ & $0.5$ & $24576$ & $10$ \\
         $30$ & -3 & $10^{-2}$ & $0.5$ & $24576$ & $10$ \\
         $20$ & -3 & $10^{-2}$ & $0.5$ & $49152$ & $10$ \\
         $30$ & -3 & $10^{-2}$ & $0.5$ & $49152$ & $10$ \\\hline
         
         $1$ & 0  & $1$ & $0.005$ & $24576$ & $0.5$ \\
         $1$ & -3 & $1$ & $0.005$ & $24576$ & $0.5$ \\
         $20$ & -3 & $1$ & $0.005$ & $24576$ & $0.5$ \\
         $30$ & -3 & $1$ & $0.005$ & $24576$ & $0.5$ \\\hline
    \end{tabular}
    \caption{Sets of parameters in our models.
    See Eq.\,\ref{eq:matter_grad} for the definitions of $\lambda_0/\mu_0$ and $m$ related to the matter gradient.
    $\xi$ denotes the attenuation parameter for neutrino oscillation scale.
    $\Delta R$ is the width of spatial domain (spherical shell) uniformly divided into $N_r$ radial cells.
    }
    \label{tab:models}
\end{table}

If the wave frequency $k_{\mu} = (\omega,\boldsymbol{k})$ has an imaginary part, the flavor coherence exponentially grows or damps in time ($\omega$) or space ($\boldsymbol{k}$).
Here, the wave frequency is shifted by the diagonal matter and neutrino potentials as backgrounds.
Note that in the (shifted) co-rotating frame, the dispersion relation is not modified by local lepton density and the gradient.
In returning to the true frame $K_{\mu} = k_{\mu} + \xi\left(\Lambda_{\mu}^{ex} + \Phi^{ex}_{\mu}\right)$ with $\Phi_{\mu}^{ex} \equiv \int\dd\Gamma\,v_{\mu}(\rho_{ee}-\rho_{xx})$, the impact of background potentials appears on the dispersion relation, but the growth rate remains unchange irrespective of the choice of the frame.
Following the dispersion relation on the true frame in the linear phase, the off-diagonal flavor coherence evolves with time and space on the plane-wave ansatz:
\begin{equation}
    \rho_{\nu}^{ex} \propto \tilde{Q}\exp\left[-\ii\Omega t +\ii Kr\right].
\end{equation}
Note that the attenuation parameter only stretches the wave frequency $k_{\sigma}$ in the denominator of Eq.\,\eqref{eq:DR}.
The unstable branch is therefore preserved when $(\Omega,K)$ are measured in units of the attenuated self-interaction scale $\xi\mu$.

\subsection{Background Model}\label{Sec:Model}
To investigate the impact of matter inhomogeneity for FFI in global geometry, we parametrize the radial profile as follows:
\begin{equation}
    \lambda(r) = \lambda_0 \left(\frac{r}{R_{\mathrm{in}}}\right)^{m},
    \label{eq:matter_grad}
\end{equation}
where $\lambda_0$ denotes the matter density at the inner boundary $r=R_{\mathrm{in}}$ and $m$ is the power with respect to the radius.
The adopted matter gradient parameters $\lambda_0$ and $m$ are listed on Table\,\ref{tab:models}.
The matter density $\lambda_0$ is particularly scaled with the self-interaction potential $\mu_0$ at the inner boundary.
Then, the phase shift term by the matter potential is
\begin{equation}
    \Lambda_0 = \lambda(r).
\end{equation}
The spatial components $\Lambda_{j=1,2,3}$ arise from the fluid velocity field, which is negligible compared to the neutrino velocity field.
Throughout this study, we neglect the bulk fluid velocity and assume a static matter background, so that $\Lambda_j = 0$.
Accordingly, matter inhomogeneity influences only the temporal (zeroth) component in the phase shift, whereas the spatial phase shift is determined solely by the neutrino potential.

We then adopt the Dirichlet boundary condition for incoming neutrinos, whose angular distributions for each species are given at the boundary
\begin{equation}
    g_{\nu_{\alpha}}(v) = n_{\nu_\alpha}
    \begin{cases}
        \left[ 1 + \beta_{\nu_\alpha}(v-0.5) \right] &(\mathrm{for}\,\, v \geq 0) \\
        \eta &(\mathrm{for}\,\, v < 0),
    \end{cases}
    \label{eq:ELN_boundary}
\end{equation}
where $\eta=10^{-6}$ is a dilute flux, and for outgoing neutrinos, the free boundary condition is adopted.
In our model, the neutrino number density at the inner boundary is set to $n_{\nu_e} = n_{\bar{\nu}_e} = 6\times 10^{32}\mathrm{\,cm^{-3}}$, and the anisotropic parameters as $\beta_{\nu_e} = 0$ and $\beta_{\bar{\nu}_e} = 1$.
Since we focus only on FFI, heavy-leptonic flavors $\nu_X$ do not contribute, and thereby we set $n_{\nu_X} = 0$ for the incoming components for simplicity.
Also, FFI is insensitive to the neutrino energy, so we assume a monochromatic distribution with a representative energy $E_{\nu} = 12\mathrm{\,MeV}$.
The perturbation in flavor coherence is characterized by the vacuum frequency $\omega_V = \Delta m^2\big/2E_{\nu}$, where $\Delta m^2 = 2.53\times 10^{-3}\mathrm{\,eV^2}$ and a mixing angle $\sin^2\theta_{V} = 0.0216$ in vacuum.
In our setup, the mixing angle is suppressed through the dense background matter $\lambda$.

To solve numerically Eq.\,\eqref{eq:QKE}, we newly developed ``{\tt GANTS-QK}'' ({\bf G}PU-{\bf A}ccelerated {\bf N}eutrino {\bf T}ransport {\bf S}imulator with {\bf Q}uantum {\bf K}inetics), which runs on GPU architectures.
See the details in Appendix\,\ref{Sec:App}.
To treat global geometry in the transport calculation, we adopt the attenuation parameter as $\xi = 10^{-4}, 4\times 10^{-4}, 10^{-2}$, and $1$.
We cover the spatial domain of $50\mathrm{\,km}\leq r \leq 100\mathrm{\,km}$ for $\xi = 10^{-4}$ and $4\times 10^{-4}$, whereas in cases with weaker attenuation parameters $10^{-2}$ and $1$, the spatial domain is reduced to $\Delta R=0.5\mathrm{\,km}$ and $0.005\mathrm{\,km}$ to reduce the computational complexity, respectively.
Note that the spatial range scaled by the attenuated self-interaction potential $\xi\mu_0$ at the inner boundary is identical for all cases of attenuation parameters except for $\xi = 4\times 10^{-4}$.

As in our previous studies \cite{Nagakura:2022a,Nagakura:2023a}, given the angular distributions at both boundaries, we first run the classical neutrino transport turning off the oscillation terms until the system reaches a steady state $t_{\mathrm{cls}}$.
Then, we compute the full QKEs in Eq.\,\eqref{eq:QKE} using the obtained neutrino field as an initial condition until the system reaches a quasi-steady state $t_{\mathrm{fin}}$.
Note that to reach a (quasi-)steady state in both classical and quantum transport, the integration time must exceed the flight time of nearly tangentially emitted neutrinos ($v=0$) across the shell.
Due to the global spherical geometry, such trajectories have a path length of $\sim 0.7\mathrm{\,km}$ within the shell even when the radial shell width is only $\Delta R = 0.005\mathrm{\,km}$.

\begin{figure}[t]
    \centering
    \includegraphics[width=1.\linewidth]{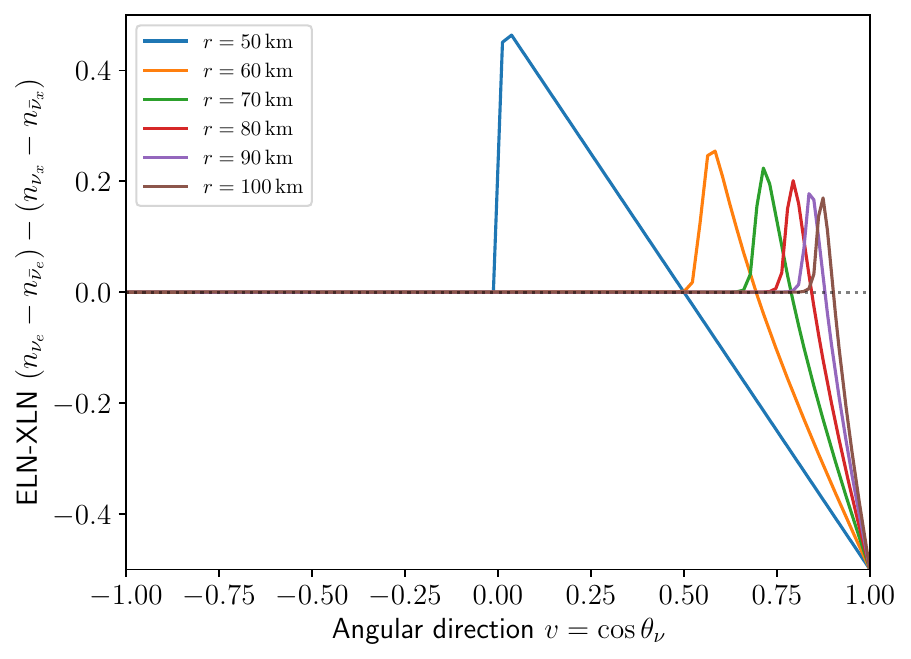}
    \caption{Radial evolution of ELN-XLN angular distributions.
    Angular distribution becomes more forward-peaked with radii due to the momentum advection.}
    \label{fig:ELN_radii}
\end{figure}
Figure\,\ref{fig:ELN_radii} shows the radial variation of ELN-XLN angular distributions in the classical steady state $t_{\mathrm{cls}}$.
Angular crossing appears at $v=0.5$ at the inner boundary $r=50\mathrm{\,km}$, as can be seen from Eqs.\,\eqref{eq:ELN_boundary}.
The momentum advection in the global spherical geometry sharpens a neutrino angular distribution injected from inner boundary, and the angular crossing is also gradually forward-peaked with propagation.
This implies that the growth rates of FFI decrease with radii due to the narrower crossings.
In the following section, the radial profile of background matter is further taken into account in the global neutrino transport with quantum kinetics.

\begin{figure*}[t]
    \centering
    \includegraphics[width=1.\linewidth]{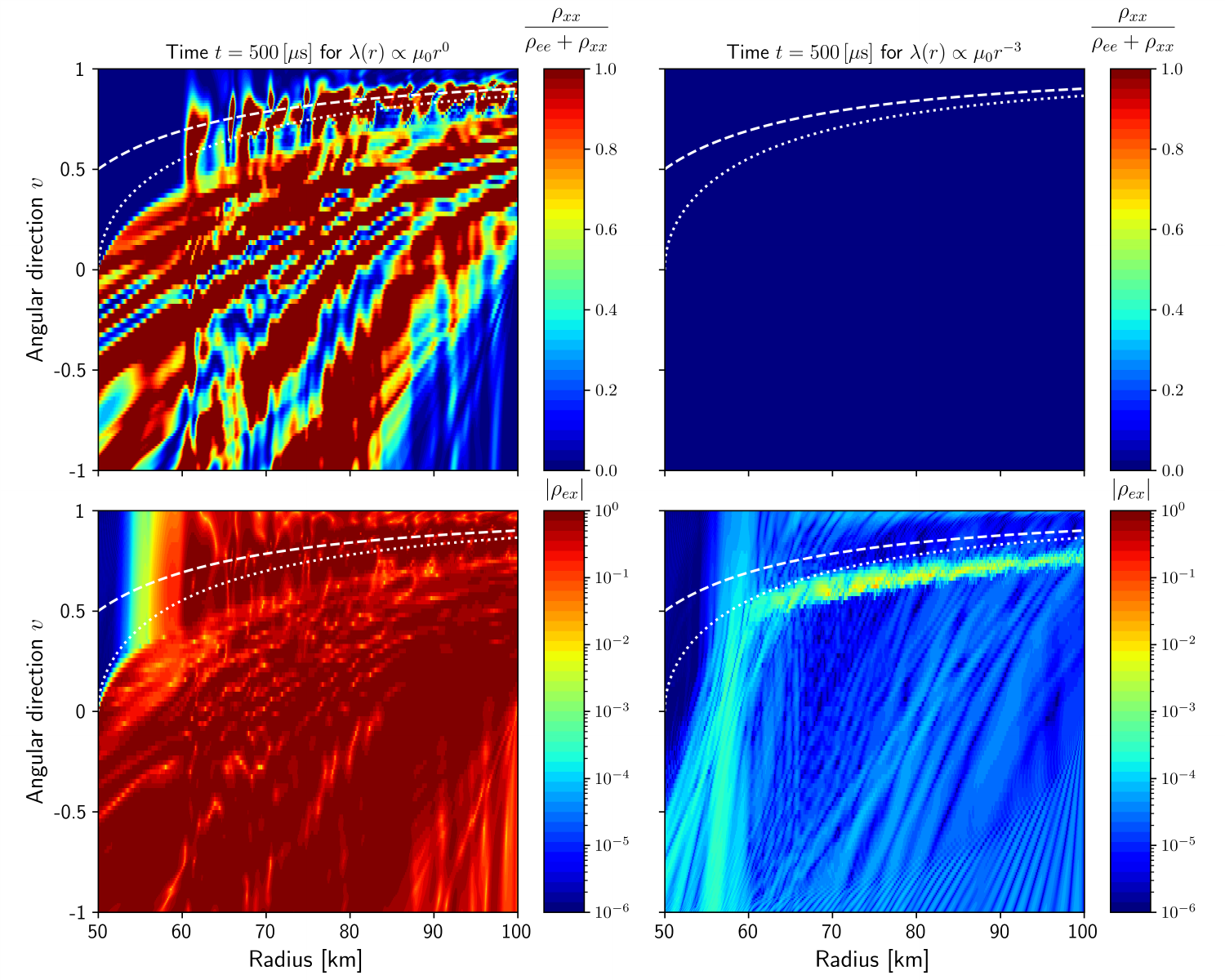}
    \caption{Radial profile of angular distributions of transition probability (top panels) and flavor coherence (bottom) at the simulation end $t_{\mathrm{fin}}$.
    These models correspond to $(\lambda_0/\mu_0, m) = (1,0)$ (left) and $(1,-3)$ (right) with attenuation $\xi = 10^{-4}$.
    White dashed and dotted lines are trajectories with $v = 0.5$ and $0$ at the inner boundary.
    See Ref.\,\cite{Zaizen2026_Movie_global} for a movie showing the full time evolution corresponding to these snapshots.
    }
    \label{fig:asympt_num}
\end{figure*}
\section{Results}
Here, we consider a radially varying background matter profile and investigate the impact on neutrino transport with FFC.
The model parameters for the background gradient are listed in Table\,\ref{tab:models}.
First, we numerically demonstrate how flavor conversion is either suppressed or occurs in the presence of matter inhomogeneity.
Then, we analytically provide the physical interpretation based on a local linear stability analysis.

\subsection{Nonlinear Flavor Conversion}\label{Sec:3A}
Figure\,\ref{fig:asympt_num} displays the radial profile of the angular distributions of the transition probability (top panels) and flavor coherence (bottom) in the case of strong attenuation $\xi=10^{-4}$ at the end of the simulations.
While the model (left) with a flat matter profile $m=0$ exhibits strong flavor conversion in the entire domain, the flavor evolution is completely suppressed in the model (right) with a steep matter gradient $m=-3$.
In particular, in the bottom panels, the flavor correlation fails to construct a coherent wave pattern over the whole angular directions around a radius of $r\sim 60\mathrm{\,km}$.
The radius corresponds to the onset of flavor conversion in the left panels, and it indicates that the presence of the background variation prevents a flavor wave from growing exponentially until reaching a linear saturation.

\begin{figure*}[t]
    \centering
    \begin{minipage}{0.45\linewidth}
        \includegraphics[width=1.\linewidth]{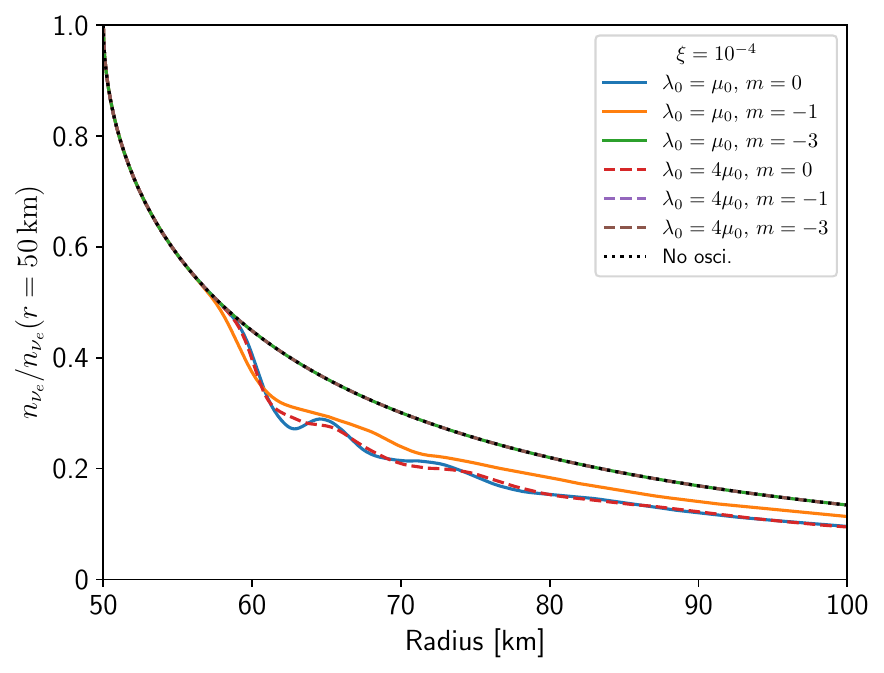}
        \includegraphics[width=1.\linewidth]{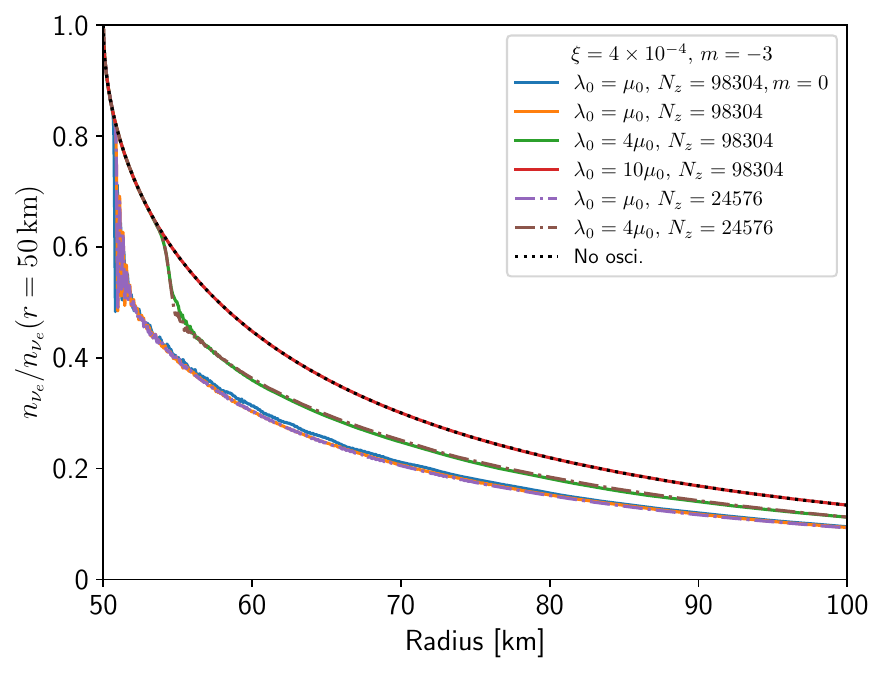}
    \end{minipage}
    \begin{minipage}{0.45\linewidth}
        \includegraphics[width=1.\linewidth]{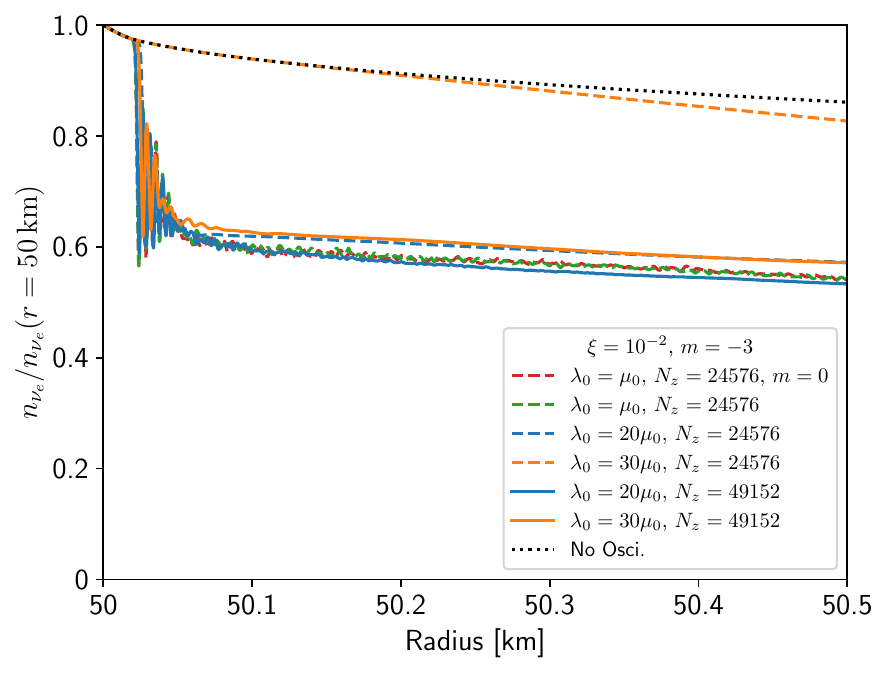}
        \includegraphics[width=1.\linewidth]{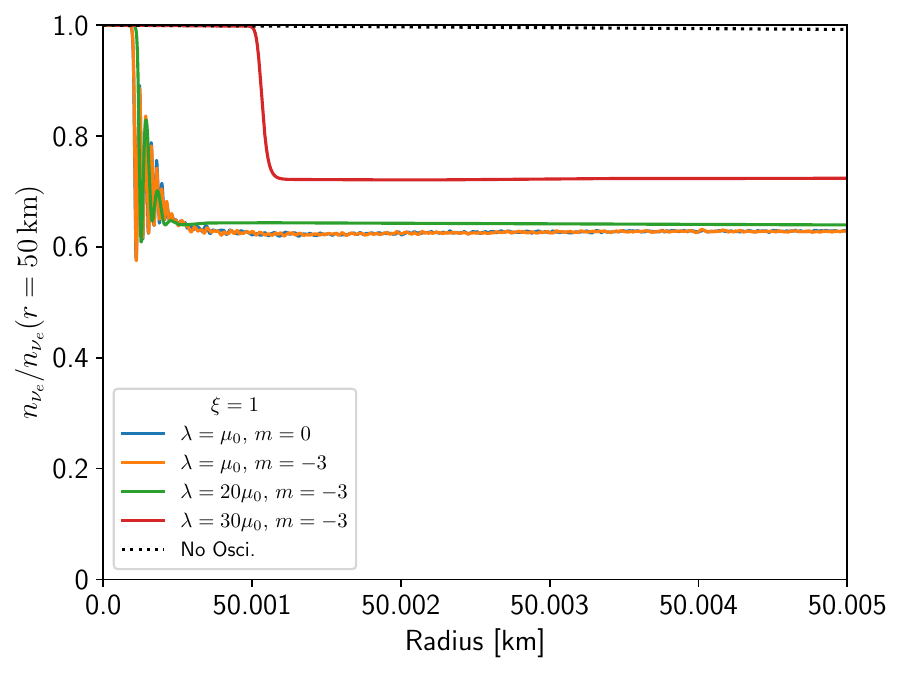}
    \end{minipage}
    \caption{Radial profile of number density of electron neutrinos time-averaged over quasi-steady states in the cases with background matter profile shown in Tab.\,\ref{tab:models}.
    Each panel corresponds to a attenuation parameter $\xi = 10^{-4}, 4\times 10^{-4}, 10^{-2}$, and $1$.
    }
    \label{fig:num_time_avg}
\end{figure*}
Figure\,\ref{fig:num_time_avg} exhibits the radial profile of electron neutrino number density time-averaged over the quasi-steady states for our models listed in Table\,\ref{tab:models}.
The degree of suppression of FFC is highly sensitive to the attenuation parameters $\xi$.

For strong attenuation models with $\xi=\mathcal{O}(10^{-4})$ (left panels), the background variation $(m\neq 0)$ hinders the growth of FFI.
On the other hand, in the bottom left panel, the models with $\lambda_0/\mu_0 = 4$ demonstrate delayed and marginal flavor conversion compared to those with lower background gradient.
Note that the behavior is consistent with the simulation result for lower resolution of $N_r = 24576$.
The background model exhibits the suppression of FFI in the case of attenuation $\xi = 10^{-4}$ (upper left) and clearly demonstrates the attenuation dependency of the effects.
The significant difference among such simulation models is the onset radius of flavor conversion.
Strong attenuation slows down the growth rate of flavor instability and increases the required propagation distance.
Consequently, the background variation that neutrinos undergo before reaching the linear saturation increases in considering strong attenuation, and flavor conversion becomes easier to be suppressed.

The behaviors can be confirmed from the weak attenuation models (upper right panel in Fig.\,\ref{fig:num_time_avg}), where powerful flavor conversion occurs for all background gradient parameters on higher spatial resolution $N_r = 49152$.
The onset radius is then $r < 50.1\mathrm{\,km}$, and the impact of the radial variation of the background is smaller than in the strongly attenuated models.
Thereby, FFI can sufficiently grow even in the higher background matter gradient.
Note that the comparison between higher resolution models (solid lines) with $N_r = 49152$ and lower ones (dashed) with $N_r = 24576$ clarifies that artificial suppression appears.
The behaviors can be understood from the unstable branches by linear stability analysis.
We will discuss it in the subsequent section.

Finally, we demonstrate no attenuation models $\xi=1$ in the bottom right panel.
Due to the small spatial domain, the radial variation of the background number density is negligible, and the flavor evolution is almost similar to the local calculations with the Dirichlet boundary \cite{Zaizen:2023a}.
Also, even for the no attenuation models, artificial matter suppression appears and the exponential growth of FFI is weakened or missed to be captured.

\subsection{Adiabaticity for Flavor Waves}\label{Sec:3B}
\begin{figure}
    \centering
    \includegraphics[width=1.\linewidth]{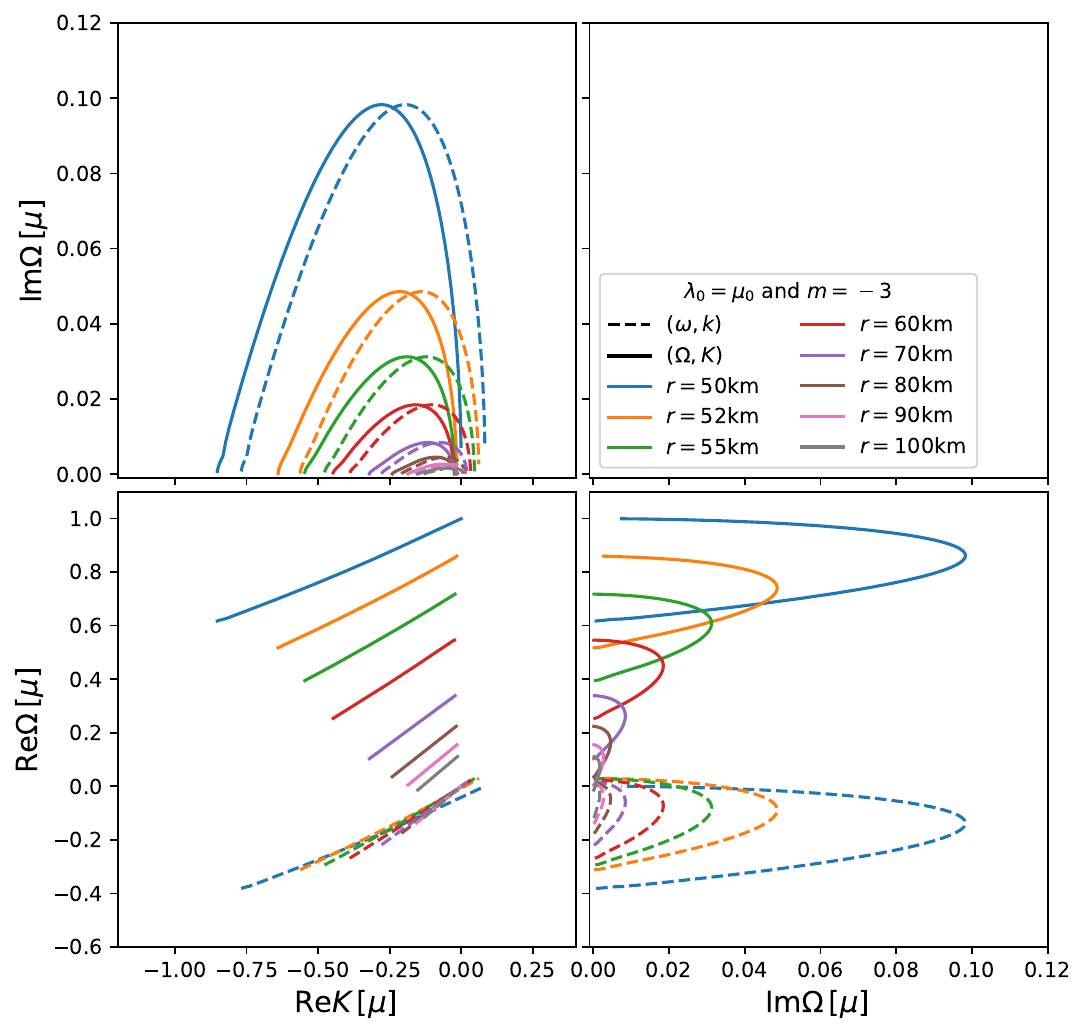}
    \caption{Dispersion relation in the case of $(\lambda_0/\mu_0,m)=(1,-3)$.
    Dashed lines are on the shifted frame $(\omega,k)$, while solid ones are on the true frame $(\Omega,K)$.
    Color contours denote dispersion relation at certain radius.
    }
    \label{fig:DR}
\end{figure}
As seen in our numerical experiments in Fig.\,\ref{fig:num_time_avg}, the suppression behavior is sensitive to the background variation, the attenuation parameter, and the spatial/temporal resolutions.
This variation in the growth of flavor coherence can be evaluated through the {\it ``adiabaticity''} of growing flavor waves.

Figure\,\ref{fig:DR} shows how the local unstable branch evolves with radius in the $\Omega-K$ plane, especially focusing on the temporal frequency $\Omega(r) = \mathfrak{R}(K(r))$ denoting the dispersion relation in terms of $K$.
As mentioned in Sec.\,\ref{Sec:Model}, the presence of background lepton shifts the unstable branch mainly along the $\mathrm{Re}\Omega$ direction (bottom right panel) rather than along the $\mathrm{Re}K$ direction (upper left).
Consequently, the unstable branch sweeps across the $\Omega-K$ plane with radial propagation and (de-)activates the corresponding flavor modes as in the left bottom panel.
Even if a flavor mode stays on the unstable branch, $\Omega_1 = \mathfrak{R}_1(K_1)$ being complex, at a certain radius $r = r_1$, it may not be possible to track the variation during propagation.
In other words, $\Omega_1 \neq \mathfrak{R}_2(K_1)$ at the other radius $r = r_2$.
Therefore, the flavor wave fails to evolve further, either being damped or just advecting without any amplification.

\begin{figure}
    \centering
    \includegraphics[width=1.\linewidth]{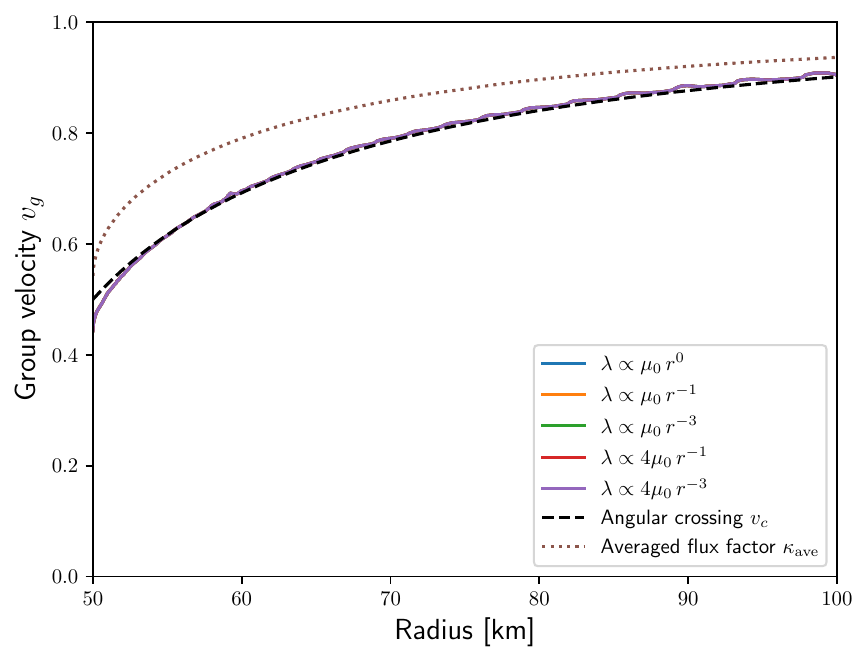}
    \caption{Radial profile of group velocity $v_g$.
    Colors indicate different matter gradients, but all of them overlap.
    Dashed and dotted lines correspond to the direction variation of angular crossing and averaged flux factor, respectively.
    }
    \label{fig:v_g}
\end{figure}From the above picture, we can define an adiabatic condition under which a flavor wave can grow sufficiently only while it remains on the local unstable branch during propagation.
As illustrated in Fig.\,\ref{fig:DR}, the unstable branch of the local dispersion relation shifts with radius as the background refractive effects vary.
This motivates a comparison between two timescales: the intrinsic growing timescale of flavor instability and the timescale over which the unstable branch drifts in the $\Omega-K$ plane during propagation.
First, the growing timescale of flavor instability is simply given by the imaginary part of the complex frequency,
\begin{equation}
    \tau_{\mathrm{inst}} \sim \left(\mathrm{Im}\Omega\right)^{-1}.
\end{equation}
Next, we estimate how rapidly the unstable branch moves as the background changes.
Since the variation is dominated by the shift along the $\mathrm{Re}\Omega$ direction as seen in Fig.\,\ref{fig:DR}, the relevant drift rate is characterized by $D_t(\mathrm{Re}\Omega)$ with $D_t$ being a Lagrangian derivative.
The corresponding timescale over which the unstable branch shifts is then
\begin{equation}
    \tau_{\mathrm{shift}} \sim \frac{\mathrm{Im}\Omega}{\abs{D_t(\mathrm{Re}\Omega)}}.
\end{equation}
Here $\mathrm{Im}\Omega$ is used not as the geometrical width of the unstable branch, but as the intrinsic frequency scale associated with one e-folding growth of the instability.
The drift timescale therefore represents the time required for the branch to move by an amount comparable to the frequency scale associated with one e-folding growth.
If $\tau_{\mathrm{inst}} \ll \tau_{\mathrm{shift}}$, the flavor wave can remain on the unstable branch long enough to experience substantial growth.
Therefore, the adiabatic condition is
\begin{equation}
\begin{split}
    &\tau_{\mathrm{inst}} \ll \tau_{\mathrm{shift}} \\
    &\Rightarrow \epsilon_{\mathrm{ad}} \equiv \frac{\abs{D_t(\mathrm{Re}\Omega)}}{(\mathrm{Im}\Omega)^2} \ll 1.
\end{split}
\end{equation}
When this condition is satisfied, neutrinos can undergo nonlinear flavor conversion.

Here, $D_t$ is a Lagrangian derivative evaluated along the propagation of the flavor wave, not as the kinetics of neutrino particles.
In the stationary background, we then write $D_t = v_g\partial_r$, where $v_g$ is the group velocity of the unstable eigenmodes.
The group velocity is $v_g = \partial \mathrm{Re}\Omega / \partial \mathrm{Re}K$, which can be read out from the bottom left panel in Fig.\,\ref{fig:DR}.
Note that the group velocity determines the propagation direction of flavor perturbation and does not depend on whether the frame is shifted.
Fig.\,\ref{fig:v_g} shows the radial profile of group velocity for the unstable branch in some background models.
Clearly, it is not sensitive to the background matter gradient and the direction is forward-peaked due to the momentum advection similar to the angular crossing.

Consequently, the adiabatic condition that flavor coherence can grow sufficiently with advection is given by
\begin{equation}
    \epsilon_{\mathrm{ad}}(r) = \frac{\Big\lvert\left[\partial_K\mathrm{Re}\Omega(r)\right] \left[\dd_r\mathrm{Re}\Omega(r)\right]\Big\rvert} {\left[\mathrm{Im}\Omega(r)\right]^{2}} \ll 1
    \label{eq:adiabaticity_1}
\end{equation}
and is determined solely by the local dispersion relation.
Since the group velocity $v_g$ and the growth rate $\mathrm{Im}\Omega$ are insensitive to the phase shift due to the background, the impact of matter gradient appears only in the radial derivative part $\dd_r \mathrm{Re}\Omega$.

Note that this adiabaticity Eq.\,\eqref{eq:adiabaticity_1} clarifies the $\xi$ dependence.
Under attenuation, oscillation frequency $(\Omega,K)$ is scaled with the attenuated self-interaction strength $\xi\mu$, whereas the group velocity $v_g = \partial_K\mathrm{Re}\Omega$ remains nearly unchanged.
Consequently, the gradient $\partial_r\mathrm{Re}\Omega$ carries one power of $\xi$, while $\left[\mathrm{Im}\Omega\right]^2$ in the denominator carries two, leading to the relation:
\begin{equation}
    \epsilon_{\mathrm{ad}}(r;\xi) = \xi^{-1}\,\epsilon_{\mathrm{ad}}(r),
\end{equation}
where $\epsilon_{\mathrm{ad}}(r;\xi)$ denotes the adiabaticity employing the attenuation.
Physically, attenuation slows the local growth rate of flavor instability, while the radial variation scale of the background remains unchanged.
As a result, it becomes more difficult to track the unstable branch adiabatically for smaller $\xi$.
Since strong attenuation is required to resolve the global geometry, the adiabatic condition $\epsilon_{\mathrm{ad}}(r;\xi) \ll 1$ can be easily violated due to the increase.

Before applying the adiabaticity to our numerical demonstrations, it is worthwhile to compare our local stability analysis to the comoving-frame approach by Ref.\,\cite{Bhattacharyya:2025}.
We can also derive the adiabaticity for a flavor wave by considering only the temporal frequency $\Omega \in \mathbb{C}$ and ignoring the spatial mode $K$.
Focusing on the maximum growing mode in Fig.\,\ref{fig:DR}, the corresponding frequency mode becomes stable when it gets outside the unstable branch.
Therefore, the survival timescale can be evaluated by
\begin{equation}
    \tau_{\mathrm{shift}} \sim \frac{\Delta\Omega_{\mathrm{Re}}}{\abs{D_t\left(\mathrm{Re}\Omega\right)}},
\end{equation}
where $\Delta\Omega_{\mathrm{Re}}$ denotes the width of the unstable branch.
Therefore, the condition where a flavor wave can sufficiently evolve is that the e-folding time is much shorter than the phase shift scale:
\begin{equation}
    \tau_{\mathrm{inst}} \ll \tau_{\mathrm{shift}}.
\end{equation}
And then, the adiabaticity is recast into 
\begin{equation}
    \epsilon_{\mathrm{ad}}(r) =\frac{\Big\lvert\left[\partial_K\mathrm{Re}\Omega(r)\right] \left[\dd_r\mathrm{Re}\Omega(r)\right]\Big\rvert}{\Delta\Omega_{\mathrm{Re}}(r)\cdot\mathrm{Im}\Omega(r)}.
    \label{eq:adiabaticity_2}
\end{equation}
This measure has a structure similar to the evaluation in the previous study \cite{Bhattacharyya:2025}.

The difference between ours and their approach is the treatment for the survival timescale appearing in the denominator.
The consideration of the width of the unstable branch against our local stability analysis means that we assume the dispersion relation $\Omega_1 = \mathfrak{R}_2(K_1)$ is satisfied between different radii $r_1$ and $r_2$ even during the propagation.
As illustrated in Fig.\,\ref{fig:DR}, this picture is, however, incorrect with respect to the dispersion relation involving radial variation.
On the other hand, Ref.\,\cite{Bhattacharyya:2025} considered the time evolution of the eigenmodes on the comoving frame along the matter inhomogeneity $\lambda[z]$ with propagation of neutrino beam.
This approach can exhibit the contribution of matter gradient to spatial Fourier mode $k$ directly and demonstrate the excellent agreement with the suppression behaviors.
However, it is not what explicitly provides the dispersion relation $\Omega = \mathfrak{R}(K)$ and is not completely identical to the variation in the $\Omega-K$ plane as described in our approach.
From the above discussion, each adiabaticity measure cannot be applied into each stability analysis framework.

\begin{figure}
    \centering
    \includegraphics[width=1.\linewidth]{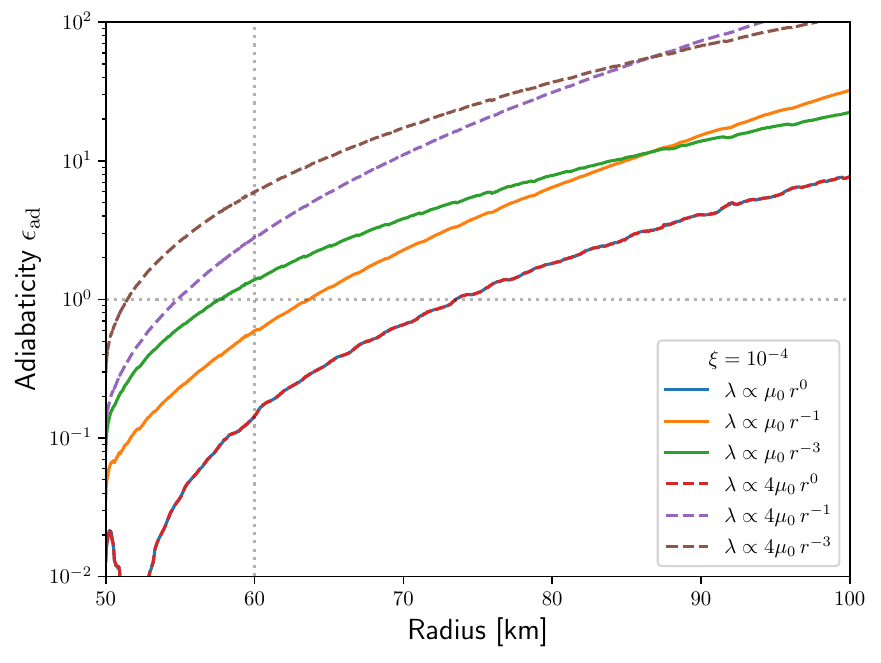}
    \includegraphics[width=1.\linewidth]{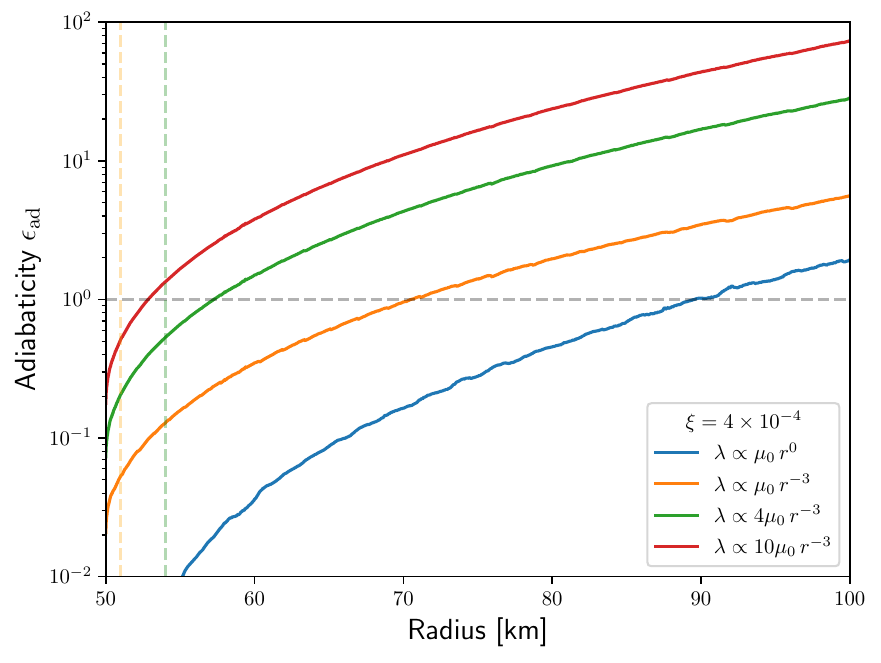}
    \includegraphics[width=1.\linewidth]{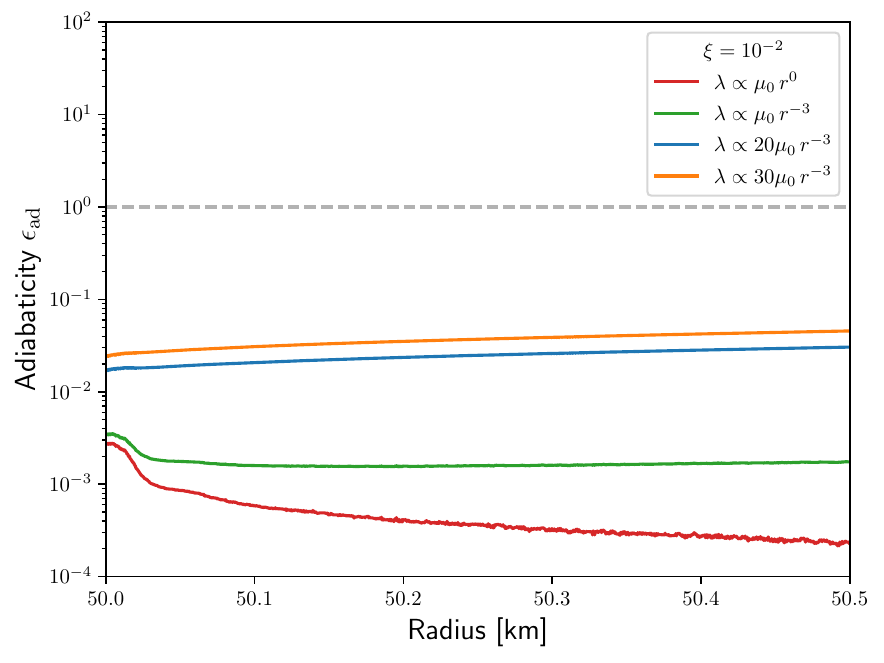}
    \caption{Radial profile of adiabatic condition $\epsilon_{\mathrm{ad}}$ for the cases of $\xi = 10^{-4}$ (top), $4\times 10^{-4}$ (middle) and $10^{-2}$ (bottom).
    Colors correspond to the difference in matter gradient parameter set $\{\lambda_0, m\}$.
    Vertical lines show the onset radii of flavor conversion in Fig.\,\ref{fig:num_time_avg}.
    }
    \label{fig:adiabatic}
\end{figure}
Figure\,\ref{fig:adiabatic} shows the radial variation of the adiabaticity in Eq.\,\eqref{eq:adiabaticity_1} for the cases of $\xi = 10^{-4}, 4\times 10^{-4}$ and $10^{-2}$.
The two upper panels clarify that the background lepton gradient can significantly distort the adiabaticity against the flavor evolution.
The behaviors are consistent with our results for $\xi = 10^{-4}$ in the upper left panel of Fig.\,\ref{fig:num_time_avg}.
The onset radius of flavor conversion is around $60\mathrm{\,km}$ (with vertical dotted line) and the adiabatic condition is satisfied for the model with flat background matter $(\lambda_0/\mu_0,m) = (1,0)$.
Also, the models with $(\lambda_0/\mu_0,m) = (1,-1)$ and $(4,0)$ exhibit the adiabatic flavor evolution in both Figs.\,\ref{fig:num_time_avg} and \ref{fig:adiabatic}.
For the other models, the adiabatic condition is violated before the onset radius and exhibits that the exponential growth of flavor waves is defeated by the background variation.

\begin{figure}
    \centering
    \includegraphics[width=1.\linewidth]{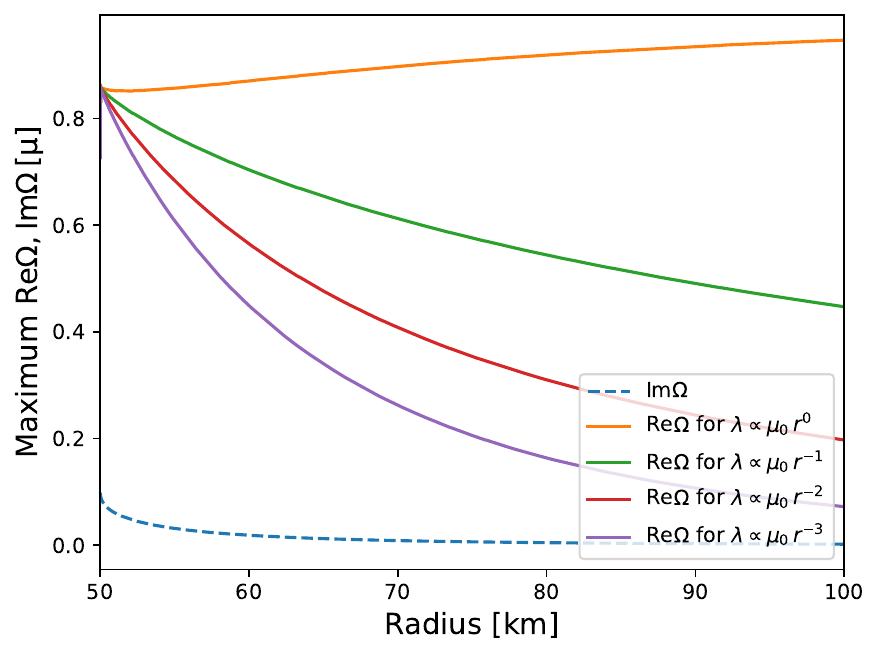}
    \caption{Radial profile of the maximum growth rate $\mathrm{Im}\Omega$ and the corresponding frequency $\mathrm{Re}\Omega$ for $\lambda_0 = \mu_0$.
    Even if the background matter profile is flat $m=0$, the advection that brings forward-peaked neutrino angular distributions alters the dispersion relation with propagation.
    }
    \label{fig:max_DR}
\end{figure}
Interestingly, even in the case of flat matter profile $m=0$, the adiabaticity can break down at large radii.
The global spherical geometry decreases the neutrino number density and brings more forward-peaked angular distributions with radius.
Thereby, the neutrino potential $\Phi_0(r)$ leaves non-zero $D_t(\mathrm{Re}\Omega)$ in the numerator as can be seen from Fig.\,\ref{fig:max_DR}, and furthermore, the sharper angular crossing at large radii has a weaker flavor instability.
Consequently, the global advection brings the violation $\epsilon_{\mathrm{ad}}>1$ of the adiabaticity at large radii even in the case of $m=0$.

In the case of attenuation $\xi=4\times 10^{-4}$, the same behaviors are seen, while the adiabaticity appears correspondingly lower due to the milder attenuation.
And compared to the flat matter case, the adiabaticity with $(\lambda_0/\mu_0,m) = (1,-3)$ is much below unity inside the onset radius.
However, in the models with $\lambda_0 \geq 4\mu_0$, the adiabaticity is relatively high even at the inner boundary so that flavor waves cannot sufficiently grow with propagation.
The onset is delayed and incomplete flavor conversion occurs in the case of $\lambda_0 = 4\mu_0$.

And in the case of weak attenuation $\xi=10^{-2}$, the adiabaticity is less enhanced and satisfies the condition even at larger radii.
As a consequence, flavor waves can lead to nonlinear flavor conversion in all of our setups.
This result also clarifies that the suppression appearing under higher matter density models is not related to the adiabatic condition and is artificial.

\subsection{Approximate Formula for Adiabaticity}\label{Sec:3C}
It is necessary to perform a linear stability analysis directly using complete momentum-space information to evaluate the adiabaticity of flavor conversion.
This approach is not suitable for discussing the effects of matter inhomogeneity in dynamical CCSN and BNSM simulations; it would be preferable to have some analytical scheme, as with the BGK framework \cite{Nagakura:2024a}, which incorporates flavor conversion.
Here, we propose a quick diagnostic for the adiabaticity, given without linear stability analysis.
To this end, one needs to analytically provide the three quantities appearing in Eq.\,\eqref{eq:adiabaticity_1}.

It is well known that the growth rate of FFI can be empirically estimated as the extension from the two-beam model \cite{Morinaga:2020,Nagakura:2019}:
\begin{equation}
    \mathrm{Im}\Omega \sim \abs{\left(\int_{G_v^{ex}>0}\frac{\dd\boldsymbol{v}}{4\pi}G^{ex}_{v}\right) \left(\int_{G_v^{ex}<0}\frac{\dd\boldsymbol{v}}{4\pi}G^{ex}_{v}\right)}^{1/2},
\end{equation}
where $G^{ex}_{v}$ denotes ELN-XLN angular distribution:
\begin{equation}
    G^{ex}_{v} = \sqrt{2}\GF\int\frac{E_{\nu}^2\dd E_{\nu}}{2\pi^2}\left[\left(f_{\nu_e} - f_{\bar{\nu}_e}\right) - \left(f_{\nu_x} - f_{\bar{\nu}_x}\right)\right]
\end{equation}
with a neutrino occupation distribution $f_{\nu_\alpha}$ specified by a flavor $\alpha$.

Obtaining the group velocity $v_g$ directly is not in the easy way as well, and the analytical estimation has not been found yet.
So, we instead choose either angular crossing direction $v_c$ or flavor-averaged flux factor $\kappa_{\mathrm{ave}}(\times c)$ as another representative velocity.
The averaged flux factor is here defined as
\begin{equation}
    \kappa_{\mathrm{ave}} = \frac{F_{\nu_e} + F_{\bar{\nu}_e}}{c\left(N_{\nu_e} + N_{\bar{\nu}_e}\right)},
\end{equation}
where $F_{\nu}$ and $N_{\nu}$ denote the first and zeroth angular moments.
For heavy-leptonic flavor neutrinos $\nu_x$, the contribution to FFI should be negligible and dropped due to the small or zero XLN in a realistic environment.
Figure\,\ref{fig:v_g} compares among $v_g$, $v_c$, and $\kappa_{\mathrm{ave}}(\times c)$ with radius.
All quantities can capture the forward-peaked behaviors, and the disparities are not large.
It should be noted that, though the angular crossing direction appears to closely follow the group velocity in our models, it is not a universal phenomenon.
For instance, in the case of preshock region of CCSNe, the group velocity is much faster than the crossing direction and it would not be very adequate as a representative \cite{Morinaga:2020}.
From this point, the flavor-averaged flux factor is better, with also being easily applicable from the hydrodynamical simulation data.

Finally, the refraction-changing rate $\partial_r \mathrm{Re}\Omega$ is given mainly by matter profile rather than by neutrinos themselves as seen in Fig.\,\ref{fig:max_DR}.
This figure shows the radial profile of the maximum growth rate $\mathrm{Im}\Omega$ (dashed) and the corresponding frequencies $\mathrm{Re}\Omega$ (solid lines) for $\lambda_0 = \mu_0$.
In the presence of matter gradient, the frequencies with the maximum growth rate are shifted to the negative direction with radius.
On the other hand, focusing on the flat matter profile $m=0$, the frequency $\mathrm{Re}\Omega$ rather increases with radius, and it means that the neutrino background can shift the unstable branch, too.
The contribution is, however, minor compared to the matter background, and consequently we can approximate it to $\partial_r\mathrm{Re}\Omega \approx \partial_r \lambda$.
The matter inhomogeneity is given just with the electron number density $\lambda = \sqrt{2}\GF(n_{e^-} - n_{e^+})$ because the on-shell muons and tauons are negligible within the explosive phenomena.

From the above, the adiabaticity can be approximately estimated without a full linear stability analysis as
\begin{equation}
    \epsilon_{\mathrm{ad}}(r) \sim \abs{\frac{\kappa_{\mathrm{ave}}(r)\,\Big[\partial_r \lambda(r)\Big]}{\left[\int_{G_v^{ex}>0}\frac{\dd\boldsymbol{v}}{4\pi}G^{ex}_{v}(r)\right] \left[\int_{G_v^{ex}<0}\frac{\dd\boldsymbol{v}}{4\pi}G^{ex}_{v}(r)\right]}}.
\end{equation}
Since both the matter potential $\lambda$ and each partial angular integral of $G^{ex}_v$ are measured in $\mathrm{km^{-1}}$, $\partial_r\lambda$ and the denominator have units of $\mathrm{km^{-2}}$.
Therefore, with the dimensionless flavor-averaged flux factor $\kappa_{\mathrm{ave}}$, the resulting estimate for $\epsilon_{\mathrm{ad}}$ is dimensionless and can be directly compared with unity.
This analytical indicator of adiabaticity is only a rough estimate, but the discrepancy is likely to remain within a factor of a few rather than an order of magnitude.
Although this approximate measure does not replace a full linear stability analysis, it provides a practical estimate of whether background variation is likely to prevent sustained growth of flavor waves.
Since it can be evaluated directly from background quantities available in classical neutrino transport calculations, it may serve as a useful diagnostic for CCSN and BNSM simulations.

\subsection{Suppression Induced by Finite Resolution}
\begin{figure}
    \centering
    \includegraphics[width=1.\linewidth]{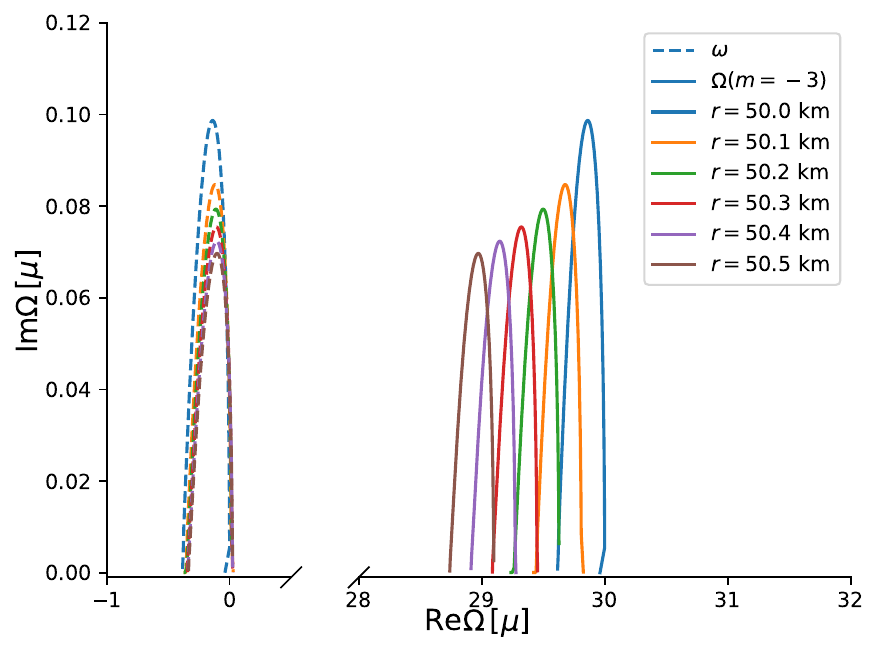}
    \caption{Dispersion relation with $(\lambda_0/\mu_0,m) = (30,-3)$ and $\xi=10^{-2}$.
    }
    \label{fig:DR_lam30}
\end{figure}
Finally, we discuss the artificial suppression appearing in the models with $\xi=10^{-2}$ in Fig.\,\ref{fig:num_time_avg}.
In the present setup, the dominant numerical limitation is the temporal resolution set by the CFL time step, although it is improved indirectly by increasing $N_r$.
Figure\,\ref{fig:DR_lam30} exhibits that the unstable branch moves away to much smaller scale (higher frequency) because of the large background matter potential.
If the time step is too large to resolve these frequencies, the exponentially growing modes are under-resolved and flavor conversion is artificially suppressed \cite{Nagakura:2025}.

The maximum resolvable frequency in our setup is roughly given by
\begin{align}
    \Omega_{\max} &\sim 1/\Delta t = C_{\mathrm{CFL}}^{-1} \frac{N_r}{\Delta R} \notag\\
        &\simeq 31 \,\mathrm{[\xi\mu_0]},
\end{align}
where $N_r = 24576$ and normalized with the neutrino self-interaction potential at the inner boundary.
The scale is comparable to the location of the corresponding unstable branch in Fig.\,\ref{fig:DR_lam30}, indicating that the growing modes are only marginally resolved.
This interpretation is also consistent with the recovery of flavor conversion at larger $N_r$, which reduces $\Delta t$ through the CFL condition.



\section{Summary and Discussion}\label{Sec:4}
In this study, we have presented fast neutrino-flavor conversion (FFC) in global spherical geometry with a background matter potential.
The global geometry brings the radial variations of both electron and neutrino number densities, thereby altering the refractive effects during propagation.
The numerical simulations have demonstrated that FFC can be suppressed by the radial gradient and the results are qualitatively consistent with the previous work using local simulations \cite{Bhattacharyya:2025}, while the suppression behaviors are strongly sensitive to the attenuation applied to the oscillation Hamiltonian.
We have also investigated whether flavor coherence can sufficiently grow with propagation in the context of the adiabaticity based on local linear stability analysis.

We parametrized the background matter profile and simulated the global flavor evolution for some attenuation parameters.
The appearance of FFC is then sensitive to the employed attenuation parameters and background properties.
For stronger attenuation (smaller $\xi$), the flavor conversion is more easily suppressed due to the steep density gradients.
This means that the background variation becomes more enhanced against the growing time of flavor coherence because the advection scale is not attenuated.

The behaviors can be understood through the variation in the dispersion relation.
The refractive effects acting on neutrino flavor waves continue to change during their propagation, and the wave mode undergoing a flavor instability is gradually shifted.
Thereby, the flavor wave cannot necessarily grow adiabatically as it propagates.
The adiabaticity can be defined as the competition between the growing timescale and the background variation.
From this evaluation, the condition is sensitive to the attenuation and clarifies our numerical demonstrations.

Our numerical and analytical investigations suggest that attenuation and finite resolution can artificially suppress collective neutrino oscillations.
Particularly, collisional and slow flavor instabilities, which have relatively longer growing timescale, would exhibit global rather than local growth \cite{Fiorillo:2025d}.
Such behaviors may require global quantum kinetic neutrino transport simulations with an attenuated oscillation Hamiltonian because of the enormous computational costs.
Thereby, even when flavor conversion does not appear under a global background matter profile, one should verify whether the suppression is physical or artificial.

Although we have carried out a parametric study of background properties and attenuation, initial neutrino distributions have been fixed throughout this paper.
Since the adiabaticity is based on the competition between growing timescale and refraction-changing timescale, our numerical modeling and evaluation depend on the growth rate of FFI.
Particularly, in the environments of CCSNe, the angular crossings tend to be narrower than in our setups.
To discuss the impact of FFC generally, one needs to take into account the realistic background properties.

\section{Acknowledgments}
We are grateful to Yudai Suwa and Chinami Kato for fruitful comments and discussions.
M.Z. is supported by Grant-in-Aid for JSPS KAKENHI Grant Numbers JP24H02245 and JP25K17383.
Numerical computations were carried out on the GPU cluster and XD2000 at the Center for Computational Astrophysics, National Astronomical Observatory of Japan.
H.N. is supported by Grant-in-Aid for Scientific Research (23K03468), the NINS International Research Exchange Support Program, and the HPCI System Research Project (Project ID: hp250006, hp250226, hp250166, hp260058).

\appendix
\section{GANTS-QK}\label{Sec:App}
{\tt GANTS-QK} is a quantum kinetic neutrino transport code designed for multi-GPU clusters using MPI.
This code parallelizes the neutrino density matrices $\rho_{\nu}$ over phase space $(\boldsymbol{r},\boldsymbol{p})$ across multiple GPU devices and follows the flavor evolution with time in the global geometry:
\begin{equation}
    \left(p^{\mu}\frac{\partial}{\partial x^{\mu}} - \Gamma^{j}_{\sigma\kappa}p^{\sigma}p^{\kappa}\frac{\partial}{\partial p^j}\right)\rho_{\nu} = -\ii \left[\mathcal{H}_{\mathrm{osc}},\rho_{\nu}\right] + \mathcal{C}[\rho_{\nu}].
\end{equation}
Here $\Gamma^{j}_{\sigma\kappa}$ is a connection coefficient for the global geometry, which gives a momentum advection for the spherical coordinate even on the flat spacetime.
Except for the spatial advection term, all physical quantities, that is, momentum advection, neutrino oscillation terms and incoherent collisions, are locally well-defined and do not require substantial communication among different GPU devices, provided that the spatial indices are separated on the multi-GPUs.

In the time integration, we use the fourth-order strong-stability-preserving Runge-Kutta method (SSP-RK(5,4)) \cite{Ruuth:2006,Gottlieb:2009} with a fixed time step size of $\Delta t = C_\mathrm{CFL}\,\mathrm{min}\{\Delta r, r\Delta\cos\theta_{\nu}\}$, where the Courant-Friedrichs-Lewy number $C_\mathrm{CFL} = 0.4$.

To capture small-scale structure due to turbulence-like cascade on the phase space $(\boldsymbol{x},\boldsymbol{p})$ \cite{Johns:2020a,Richers:2021,Zaizen:2021a}, the code uses a fifth-order WENO (weighted essentially non-oscillatory) finite volume scheme \cite{Jiang:1996} in evaluating the advection in coordinate and momentum space.
Also, to perform the angular integration in the neutrino self-interaction term and the collision terms more accurately, it adopts the Gauss-Legendre quadrature.
The angular bins are then distributed on the non-uniform roots of the Legendre polynomial.
Thereby, the applied WENO scheme is extended to operate on the non-uniform grids \cite{Huang:2018}.

To avoid a numerical instability due to the discontinuity, particularly in the angular distributions between forward and backward directions, we take additional limiter prescription in the WENO scheme, according to Ref.\,\cite{Nagakura:2022}.
In the WENO scheme, we usually reconstruct numerical fluxes at each cell interface, using the linear combination of interpolations on substencils.
The weight functions are
\begin{equation}
    w_j = \frac{\tilde{w_j}}{\sum_k \tilde{w}_k},\,\,\,\,\,\,
    \tilde{w}_k = \frac{\gamma_k}{(\epsilon + \beta_k)^2},
\end{equation}
with the linear weight $\gamma_k = (0.1, 0.6, 0.3)$ for the fifth-ordered WENO scheme.
Here $\beta_k$ denotes a smoothness indicator on substencil $S_k$, and $\epsilon = 10^{-6}$ is a parameter to avoid dividing by zero in the denominator.
To sustain the numerical stability more safely, we introduce a normalization factor $Q$ and another limiter $\epsilon_2$ for the original limiter $\epsilon$:
\begin{equation}
    \tilde{w}_k = \frac{\gamma_k}{Z_k},
\end{equation}
where
\begin{align}
    Z_k &\equiv \mathrm{max}\left[\epsilon_2,\, \left(\epsilon Q + \beta_k\right)^2\right] \\
    Q &\equiv \frac{\left(\sum_l\abs{\hat{f}_{j+1/2}^{(l)}}\right)^2}{9}.
\end{align}
Here, $\hat{f}_{j+1/2}^{(l)}$ corresponds to the interfacial state interpolated on substencil $S_k$.
The squared pseudo states have the same dimensional structures as the elements in the smoothness indicator $\beta_k$ allowing the limiter $\epsilon$ to act more effectively using physical quantities.
Note that both $Q$ and $\beta_k$ can be zero when the basic quantity in the transport equation is zero everywhere, so we set another limiter $\epsilon_2 = 10^{-50}$.

\bibliography{global_matter}

\end{document}